\documentclass[aps,prd,twocolumn,showpacs,floatfix,preprintnumbers,amsfont,amsmath,amssymb,nofootinbib, superscriptaddress, reprint]{revtex4-2}

\usepackage[utf8]{inputenc}
\usepackage{amsmath}
\usepackage{physics}
\usepackage{comment}
\usepackage{bm}
\usepackage{siunitx}
\usepackage{graphicx}
\usepackage{natbib}
\usepackage{mathtools}
\usepackage{xcolor}
\usepackage{ctable}
\usepackage[normalem]{ulem}
\usepackage{ctable}
\usepackage{dcolumn}

\definecolor{keynoteBlue}{HTML}{0365C0}
\usepackage[colorlinks=true,linkcolor=red,citecolor=purple, urlcolor=keynoteBlue!60!black]{hyperref}

\newcommand{\msun}{{M_\odot}}

\newcommand{\chieff}{\chi_{\rm eff}}

\newcommand{\lib}[1]{\texttt{#1}}
\newcommand{\phXPHM}{\lib{IMRPhenomXPHM}}

\newcommand{\default}{\textsc{Default}}
\newcommand{\nopre}{\textsc{No Precession}}
\newcommand{\nohm}{\textsc{No Higher Multipoles}}

\usepackage[T1]{fontenc}

\usepackage{longtable}
\usepackage{enumitem}

\begin{document}

\title{Signs of Higher Multipoles and Orbital Precession in GW151226}

\author{Horng Sheng Chia}
\email[]{hschia@ias.edu}
\affiliation{\mbox{School of Natural Sciences, Institute for Advanced Study, Princeton, NJ 08540, USA}}
\author{Seth Olsen}
\affiliation{\mbox{Department of Physics, Princeton University, Princeton, NJ 08540, USA}}
\author{Javier Roulet}
\affiliation{\mbox{Department of Physics, Princeton University, Princeton, NJ 08540, USA}}
\author{Liang Dai}
\affiliation{\mbox{Department of Physics, University of California, Berkeley, 366 LeConte Hall, Berkeley, CA 94720, USA}}
\author{Tejaswi Venumadhav}
\affiliation{\mbox{Department of Physics, University of California at Santa Barbara, Santa Barbara, CA 93106, USA}}
\affiliation{\mbox{International Centre for Theoretical Sciences, Tata Institute of Fundamental Research, Bangalore 560089, India}}
\author{Barak Zackay}
\affiliation{\mbox{Dept. of Particle Physics \& Astrophysics, Weizmann Institute of Science, Rehovot 76100, Israel}}
\author{Matias Zaldarriaga}
\affiliation{\mbox{School of Natural Sciences, Institute for Advanced Study, Princeton, NJ 08540, USA}}

\begin{abstract}

We present a reanalysis of GW151226, the second binary black hole merger discovered by the LIGO--Virgo Collaboration. Previous analysis showed that the best-fit waveform for this event corresponded to the merger of a $\sim 14 \, M_\odot$ black hole with a $\sim 7.5 \, M_\odot$ companion, and the posterior distribution in mass ratio ($q \leq 1$) is rather flat. 
In this work, we perform parameter estimation using a waveform model that includes the effects of orbital precession and higher-order radiative multipole modes, and we find that the source parameters of GW151226 shift towards the low $q$ and high effective spin ($\chieff$) region and that $q$ is better measured. The new solution has a log likelihood roughly two points higher than when either higher multipoles or orbital precession is neglected and can alter the astrophysical interpretation of GW151226. Additionally, we find it useful to use a flat-in-$\chieff$ prior, which does not penalize the large $|\chieff|$ region, in order to uncover the higher likelihood region for GW151226. Our solution has several interesting properties: (a) the secondary black hole mass is close to the upper limit of the hypothesized lower mass gap of astrophysical black hole population; and (b) orbital precession is driven by the primary black hole spin, which has a dimensionless magnitude as large as $\sim 0.85$ and is tilted away from the orbital angular momentum at an angle of $\sim 57^\circ$. Since GW151226 is a relatively weak signal, an unambiguous claim of the detection of these effects in the signal cannot be made.

\end{abstract}

\maketitle

\section{Introduction}

GW151226 marks the second confident observation of a binary black hole (BBH) merger, detected by the LIGO--Virgo Collaboration (LVC)~\cite{Abbott:2016nmj, LIGOScientific:2018mvr} on December 26, 2015. 
This event, which has a network signal-to-noise (SNR) ratio of approximately 13, is interesting because it is a low-mass binary system and was consequently observed over a large number of cycles ($\sim 55$) in the detector's sensitive frequency band. The large number of cycles allows us to characterize the signal relatively precisely and helps us draw more detailed conclusions about the underlying system. For example, one of the black holes was inferred to have non-vanishing spin along the direction of the orbital angular momentum at $\gtrsim 90\%$ confidence~\cite{Abbott:2016nmj, LIGOScientific:2018mvr, Zackay:2019tzo, Venumadhav:2019tad, Nitz:2019hdf}. 
These measurements make an important contribution to constraints on formation channels for the astrophysical BBH population~\cite{Chatterjee:2016thb, OShaughnessy:2017eks} and enable tests of general relativity in the strong-field regime~\cite{Yunes:2016jcc, LIGOScientific:2019fpa}. 

\vskip 2pt 

While the astrophysical implications of GW151226 have been extensively explored in the literature, those works relied on source parameters that were inferred using waveform models that exclude either orbital precession or higher-order radiative multipoles, or both. For instance, the parameter estimation (PE) conducted in Refs.~\cite{Zackay:2019tzo, Venumadhav:2019tad} and the base results in the official LVC papers~\cite{Abbott:2016nmj, LIGOScientific:2018mvr} utilized waveform models that assume the spins of the black holes are aligned with the orbital angular momentum, and only include the dominant quadrupolar radiation~\cite{Purrer:2015tud, Khan:2015jqa, Taracchini:2013rva}. The LVC~\cite{Abbott:2016nmj, LIGOScientific:2018mvr} and the independent PyCBC group~\cite{Nitz:2019hdf} also presented analyses with waveform models that incorporate orbital precession, but exclude higher multipoles~\cite{Hannam:2013oca, Schmidt:2014iyl, Pan:2013rra}. In contrast, Ref.~\cite{Payne:2019wmy} used the likelihood reweighting method~\cite{10.5555/1051451, 10.5555/1571802} to reanalyze GW151226 with a waveform model that includes higher multipoles but not precession~\cite{Varma:2018mmi}. The results of those studies were broadly consistent with one another, and found no significant signs of either higher multipoles or orbital precession in GW151226. 

\vskip 2pt 

Recent advancements in template waveform modeling have led to the construction of models that encapsulate \textit{both} precession and higher multipoles in a unified framework~\cite{Pratten:2020ceb, Khan:2019kot, Ossokine:2020kjp, Varma:2019csw}. In the past, these state-of-the-art models have not been used to re-analyze GW151226, or more broadly, the larger set of events identified in the first and second LVC observing runs (O1 and O2). For instance, in their population analysis paper~\cite{Abbott:2020gyp}, the LVC used the posterior samples derived with the older waveform models~\cite{Hannam:2013oca, Schmidt:2014iyl, Pan:2013rra, Taracchini:2013rva} for the O1 and O2 events, although new models which include precession and higher multipoles are used to analyze the events identified in the first half of the third observing run (O3a)~\cite{Khan:2019kot, Ossokine:2020kjp, Varma:2019csw}. Recently, we and the authors of Refs.~\cite{Mateu-Lucena:2021siq, Nitz:2021uxj} used {\phXPHM}~\cite{Pratten:2020ceb}, a state-of-the-art model which includes both higher multipoles and the effects of orbital precession in the waveform, to analyze the O1--O3a events. Interestingly, despite using the same waveform model, each group found different results in the parameter estimation for GW151226, including varying degrees of observed bimodality in the posterior distributions. These disagreements highlight an unusual difficulty in analyzing GW151226, and warrants further investigation into the nature of this interesting event. 

\vskip 2pt

In this work, we reanalyze GW151226 using {\phXPHM}~\cite{Pratten:2020ceb} and include a detailed exploration of the changes in our PE results when either higher multipoles (HM) or orbital precession is disabled in the signal model. While our base analysis is conducted using a prior that is uniform in the effective spin~\cite{Roulet:2021hcu}, $\chieff$, we also study how our results change when the commonly-used isotropic spin prior~\cite{gwtc2_LIGOScientific:2020ibl, Mateu-Lucena:2021siq, Nitz:2021uxj} is imposed (see Ref.~\cite{Olsen:2021qin} for a comparison of these priors).

\vskip 2pt

We find that the posterior distributions of the source parameters of GW151226 shift towards the extreme mass ratio and high-$\chieff$ region under the flat-in-$\chieff$ prior when both HM and orbital precession are included in the analysis (we refer this as our {\default} setup). 
In this case, our solution has a mass ratio of $q \equiv m_2/m_1 \sim 0.3$ and $\chieff \sim 0.3$, where $m_1$ and $m_2$ are the primary (heavier) and secondary (lighter) black hole (BH) masses.
On the other hand, when either HM or orbital precession is disabled, or when imposing the isotropic spin prior (which penalizes solutions in the large $|\chieff|$ region), the solutions have mass ratio $q \sim 0.45$ and $\chieff \sim 0.25$. The latter result is consistent with those reported in earlier studies~\cite{Abbott:2016nmj, LIGOScientific:2018mvr, Zackay:2019tzo, Venumadhav:2019tad, Nitz:2019hdf}, where the posterior distribution in $q$ is visibly broader than the result of our {\default} setup and strongly prefers $q > 0.25$. Crucially, the likelihood of our new extreme mass ratio solution is a factor of $\sim e^2$ larger than the maximum likelihood in the high-$q$ region, so it can have an impact on the astrophysical interpretation of GW151226. This observation highlights the importance of both HM and orbital precession in analyzing GW151226, 
though we note that GW151226's SNR of $\sim 13$ makes it a relatively weak signal to make an unambiguous claim of the detection of these effects.
While our flat-in-$\chi_{\rm eff}$ prior is not the most astrophysically probable spin prior for merging black holes, our finding demonstrates that prior assumptions on the BH spin distribution which penalize large $|\chieff|$ can result in a failure to uncover the higher likelihood region for GW151226.

\vskip 2pt

In addition to having a mass ratio far from unity and displaying signs of precession, our solution is qualitatively interesting for a variety of reasons.  
Firstly, the secondary BH mass of the new solution has a median and $90\%$ symmetric credible interval of $m_2 = 5.8^{+2.9}_{-1.7}  M_\odot$, giving the posterior non-negligible overlap with the hypothesized lower mass gap of black hole populations (roughly $2.5$--$5 M_\odot$~\cite{Bailyn:1997xt, Ozel:2010su, Farr:2010tu}). 
Secondly, the primary BH is found to have a very large spin magnitude, $|\vec{\chi}_1| = 0.85^{+0.13}_{-0.35}$.
Furthermore, we find that the primary spin is tilted away from the orbital angular momentum at $\theta_{1L} = ({57_{-23}^{+37}})^\circ$ and is therefore the main driver of precession. The secondary spin, on the other hand, is unconstrained. 
Since these source properties are different from those inferred in the literature~\cite{Abbott:2016nmj, LIGOScientific:2018mvr, Zackay:2019tzo, Venumadhav:2019tad, Nitz:2019hdf}, GW151226 could be a binary system with previously unexplored astrophysical implications. These results serve to motivate further advancements in the waveform modeling frontier, which will allow us to more accurately determine the source properties of existing and future detections of compact binary coalescence signals.

\vskip 3pt

This paper is organized as follows: in Section~\ref{sec:PEsetup} we describe our PE method and outline the different setups used to decipher the underlying physics of GW151226. In Section~\ref{sec:results} we report the parameter estimation results. In Section~\ref{sec:Discussion} we elaborate on our findings, examining how the presence of HM and orbital precession in the signal model affect the inferred source properties, such as the spin of the primary BH. In Section~\ref{sec:astro} we discuss the astrophysical implication of our new solution. Finally, we summarize and conclude in Section~\ref{sec:conclusion}.

\section{Parameter estimation setups} \label{sec:PEsetup}

We perform PE across a suite of different computational setups in order to investigate the origin of the shift towards the low-$q$ (extreme mass ratio) and high-$\chieff$ region. In all of these computations, we use the recently developed {\phXPHM} waveform approximant~\cite{Pratten:2020ceb}, which models the GW emitted by a quasi-circular precessing BBH. In addition to the dominant quadrupolar radiation, the model also includes higher-order multipoles in the co-precessing frame of the binary system (see below for the list of available multipole modes). {\phXPHM} calibrates analytic expressions of these radiative multipoles, which are accurate in the early-inspiral regime of the binary evolution, with numerical relativity simulations that describe the merger-ringdown regime. Since the simulations have only been performed for mass ratios of $0.25 < q < 1$ for precessing binaries~\cite{Varma:2019csw}, merger-ringdown regime at smaller values of $q$ necessarily requires extrapolation.\footnote{In fact, the need to extrapolate for binaries with $q < 0.25$ in the merger-ringdown regime occurs for all of the current waveform models that incorporate precession and higher multipoles~\cite{Pratten:2020ceb, Khan:2019kot, Ossokine:2020kjp}.\label{Footnote:extrapolate}} For GW151226, where a large number of orbital cycles were observed in the early-inspiral stage and the inferred value of $q \approx 0.3$ is within the bounds, the waveform model remains a good description~\cite{Varma:2019csw, Pratten:2020ceb}. To evaluate the dynamic evolution of the binary's Euler angles due to orbital precession, we use the default precession-twisting prescription of {\phXPHM}~\cite{Pratten:2020ceb}, which is a hybrid of the so-called multi-scale analysis method~\cite{Chatziioannou:2017tdw} and the post-Newtonian approximation~\cite{Marsat:2013wwa}.

\vskip 4pt

The specifications of our parameter inference setups are detailed as follows:
\begin{itemize}
    \item \default: our main setup in which we include orbital precession and all of the multipole moments in {\phXPHM}, which are the $(\ell, |m|) = \{ (2,2), (2,1), (3,3), (3,2), (4,4) \}$ multipoles in the co-precessing frame~\cite{Pratten:2020ceb};
    
    \item \nohm : same as {\default}, except we only include the dominant $(\ell, |m|) = (2, 2) $ multipole in the co-precessing frame of the binary. This is achieved through the multipole selection feature of the LALSuite~\cite{lalsuite} implementation of the waveform model;
    
    \item \nopre: same as {\default}, except we disable precession in the waveform model by setting the in-plane spin components of both BHs to zero during likelihood evaluation~\cite{Zackay:2019tzo}.\footnote{In principle, setting the in-plane spins to zero would also nullify the effects of in-plane spin-spin interactions on the phase of the gravitational waveform~\cite{Kidder:1992fr, Kidder:1995zr}. However, since the spin-spin interactions are neglected in the modelling of {\phXPHM}~\cite{Pratten:2020ceb}, our proposed test is equivalent to simply disabling precession.}

\end{itemize}
The {\nohm} and {\nopre} setups are designed to disentangle the effects of higher multipoles and orbital precession from the parameter inference under the {\default} setup. We find that only the posterior distributions of the {\default} setup have significant support in the low-$q$ and high-$\chieff$ region.

\begin{figure*}[ht]
    \centering
    \includegraphics[width=\linewidth]{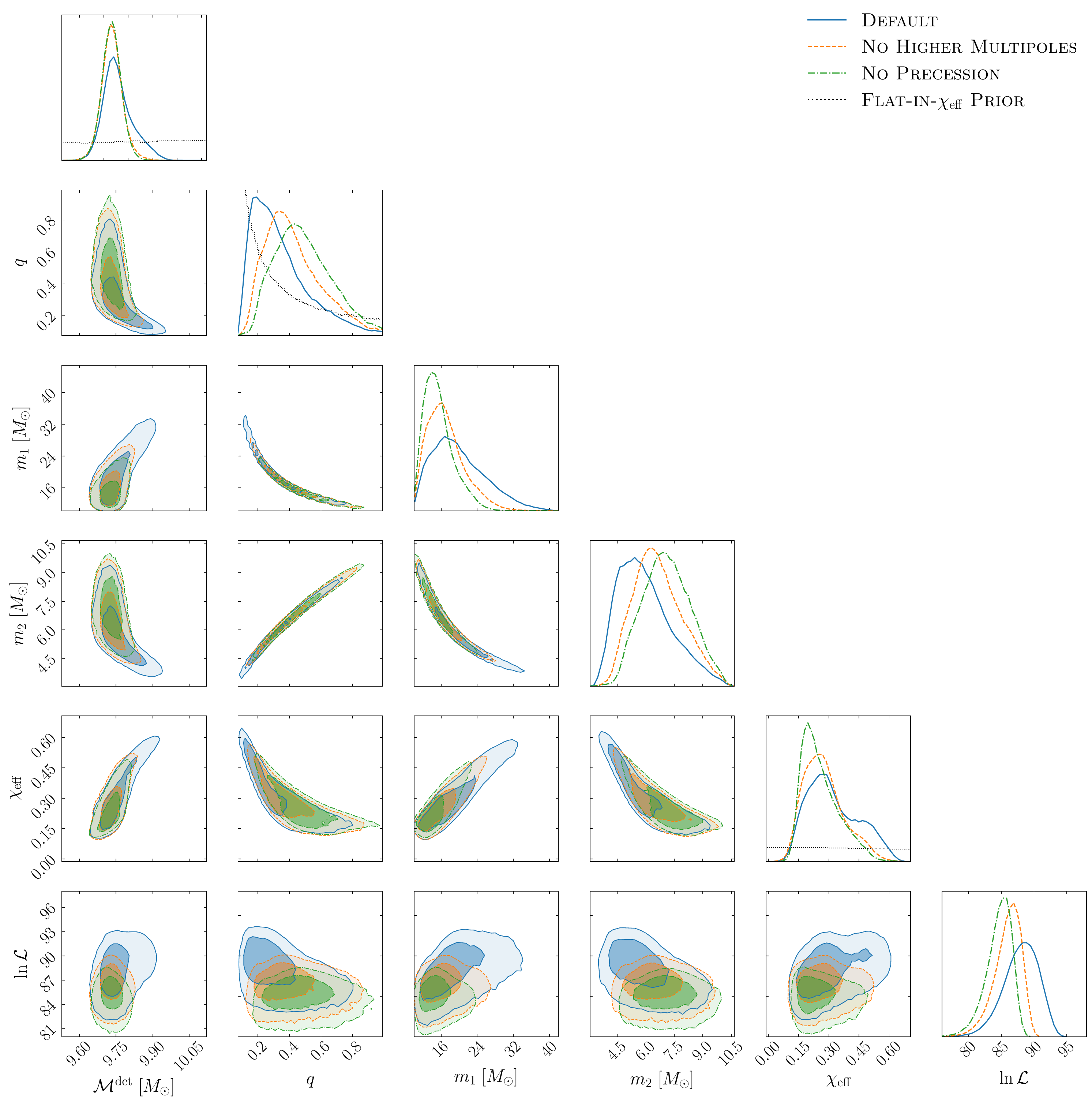}
    \setlength{\belowcaptionskip}{0pt}
    \caption{Posterior distributions for GW151226 on the detector-frame chirp mass ($\mathcal{M}^{\rm det}$), the mass ratio ($q \equiv m_2/m_1 \leq 1$), the primary BH source-frame mass ($m_1$), the secondary BH source-frame mass ($m_2$), the effective spin ($\chieff$), and the log likelihood ($\ln \mathcal{L}$). The 2--D contours enclose the 50\% and 90\% credible regions. The priors used in these setups are shown on the diagonal subplots. We see that the inclusion of HM and orbital precession in the parameter inference shifts the log likelihood and posterior distributions toward the low-$q$ and high-$\chieff$ region.}
    \label{fig:corner_plot}
\end{figure*}

\vskip 6pt

In the setups described above, we adopt a prior that is uniform in the detector-frame constituent masses, which is equivalent to a prior that is uniform in the detector-frame chirp mass $\mathcal{M}^{\rm det}$ and one that has higher prior probability for small mass ratios $q$ (see Fig.~\ref{fig:corner_plot} below). We use a flat in $\mathcal{M}^{\rm det}$ prior because the detector-frame chirp mass is the best measured mass parameter in GW astronomy. Our prior is also uniform in luminosity volume and the effective spin, $\chieff \equiv ( \vec{\chi}_{1} + q \hskip 1pt \vec{\chi}_{2}  )\cdot \hat{L} / (1+q) $~\cite{Ajith:2009bn}, where $\vec{\chi}_1$ and $\vec{\chi}_2$ are the dimensionless spin vectors of the BHs and $\hat{L} \equiv \vec{L}/ | \vec{L} |$ is the unit vector along the (Newtonian) orbital angular momentum of the binary. Similarly to $\mathcal{M}^{\rm det}$, we use the flat-in-$\chieff$ prior because $\chieff$ is the best-measured spin parameter in the GW signal of a BBH merger. For the remaining spin components, we adopt a prior that is uniform over the poorly-measured $\chi_{\rm diff} \equiv (q \vec{\chi}_1 - \vec{\chi}_2) \cdot \hat{L} / (1+q)$, conditioned on $\chieff$ and enforcing the Kerr limit on the individual spins, $|\vec{\chi}_1| \leq 1$ and $|\vec{\chi}_2| \leq 1$. $\chieff$ and $\chi_{\rm diff}$ determine the two spin components that are aligned with the orbital angular momentum, $\chi_{1z}$ and $\chi_{2z}$ . We then take the prior of the in-plane spin components of the black holes, $\chi_{ix}$ and $\chi_{iy}$ with $i = 1, 2$, to be uniformly distributed in the disk $\chi^2_{ix} + \chi^2_{iy} \leq 1 - \chi^2_{iz}$. These choices result in a prior for the effective precession parameter $\chi_p$, defined in (\ref{eqn:chip}), that is illustrated by the {\nopre} curve in Fig.~\ref{fig:Chip} (recall that the {\nopre} setup is designed such that the in-plane spins are ignored in the likelihood evaluation -- the sampled posterior distributions is therefore equivalent to the prior used for these variables).

\vskip 10pt

In order to investigate the effects of prior assumptions on our solution, we evaluate an additional setup:
\begin{itemize}
    \item \textsc{Default (isotropic spin Prior)}: same as {\default}, except we replace the flat-in-$\chieff$ prior with the isotropic spin prior.
\end{itemize}
The isotropic spin prior consists of independent constituent spin distributions that are uniform in spin magnitude and isotropic in spin orientation (the priors on other parameters are unchanged). This prior is routinely used in the literature~\cite{LIGOScientific:2018mvr, gwtc2_LIGOScientific:2020ibl, Mateu-Lucena:2021siq, Nitz:2021uxj}, and, unlike the flat-in-$\chieff$ prior, it suppresses at large values of $|\chieff|$, see Fig.~\ref{fig:corner_plot_lvc} below and e.g. Ref.~\cite{Huang:2020ysn}. For a more detailed comparison of these priors, see Section II of  Ref.~\cite{Olsen:2021qin}.

\vskip 6pt

We analyze the GW time series data with a sampling frequency of \SI{2048}{\hertz} (Nyquist frequency of \SI{1024}{\hertz}) and estimate the noise power spectral density (PSD) using Welch's method, with the specifications detailed in Ref.~\cite{Venumadhav:2019tad}. Our source parameter inference is conducted with \texttt{cogwheel}~\cite{Javier}, the same recently developed code used for PE in the IAS pipeline analysis of the O3a observing run \cite{iasO3a:2022pin}. This software implements a convenient system of coordinates and uses the relative binning method~\cite{Zackay:2018qdy} for rapid likelihood evaluation (generalized to waveforms with HM \cite{Leslie2021}). Additionally, we include a PSD drift correction~\cite{Zackay:2019kkv} in order to account for the noise's leading order (linear) deviation from stationarity around the time of the event. 

\vskip 4pt

We sample the posterior using \lib{PyMultiNest}~\cite{Feroz:2007kg, Feroz:2008xx, Pymultinest}, a nested-sampling algorithm which is designed to be capable of accurately mapping out multi-modal distributions. In order to ensure that the sampler explores all the relevant regions of the parameter space (i.e., uncovers all the likelihood peaks), we use \num{20000} live points in our computations. Importantly, the same large number of live points is used in all of the parameter estimation setups described above, such that any absence of the low-$q$ and high-$\chieff$ region reflects a genuine suppression of that part of parameter space, rather than undersampling. 

\vskip -4pt

\section{Results} \label{sec:results}

\begin{figure*}[ht]
    \centering
    \includegraphics[width=\linewidth]{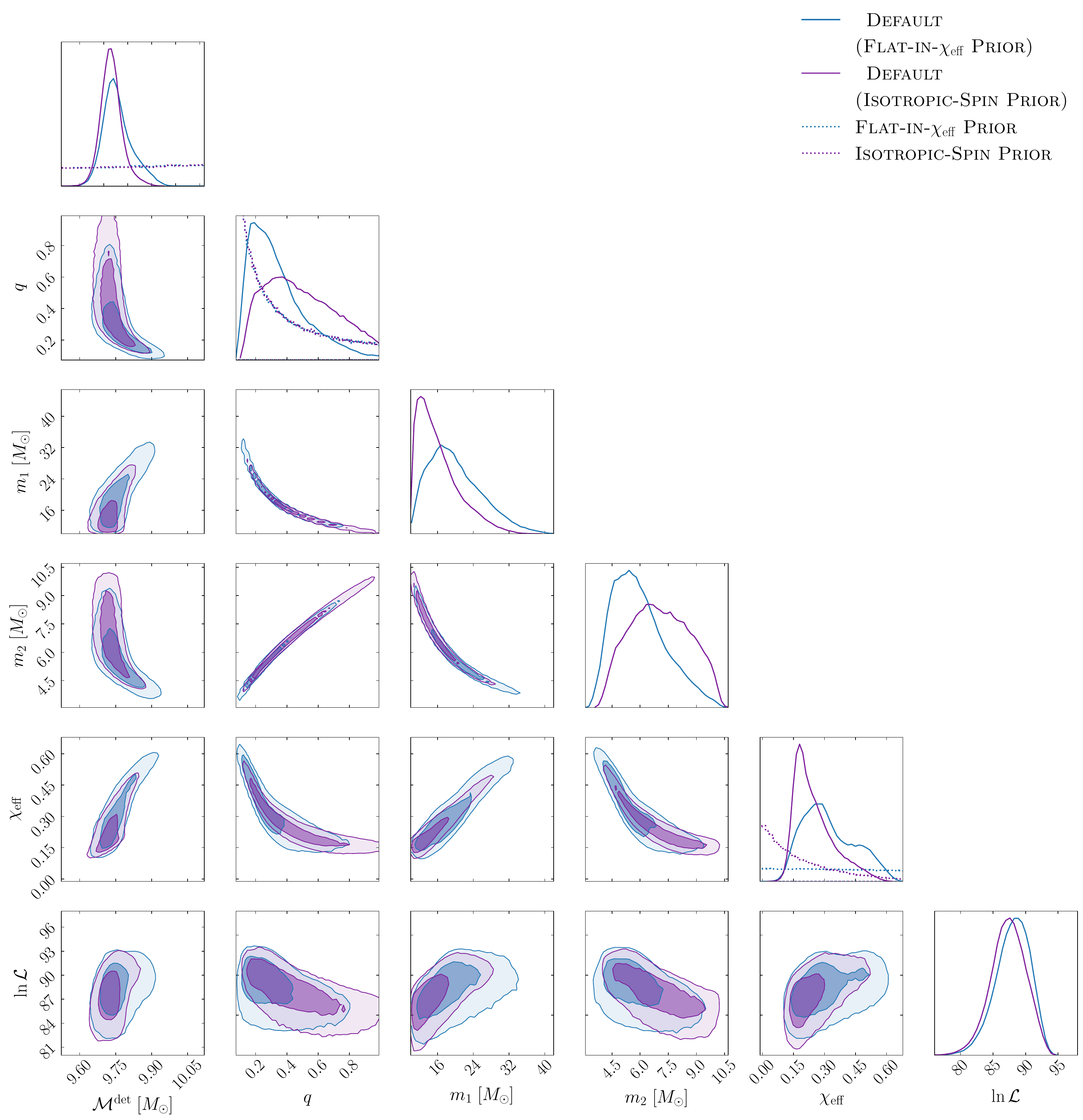}
    \setlength{\belowcaptionskip}{-10pt}
    \caption{Same as Fig.~\ref{fig:corner_plot}, except we show the posterior distributions of the {\default} setups evaluated with the flat-in-$\chieff$ prior and the isotropic spin prior. The two priors are the same apart from the spin variables. We find that the posterior distribution of $\chieff$ in the isotropic spin case is suppressed by the prior at large values of $\chieff$. Since the low-$q$ region is correlated wit the high-$\chieff$ region, the low-$q$ region is also relatively suppressed under the isotropic spin prior.}
    \label{fig:corner_plot_lvc}
\end{figure*}

In Fig.~\ref{fig:corner_plot}, we show the posterior distributions from PE under the flat-in-$\chieff$ prior. The posterior in the {\default} setup is shifted towards the low-$q$ and high-$\chieff$ region compared to the {\nohm} and {\nopre} setups. 
A summary of the inferred source parameters is presented in Table~\ref{table:params}, where we give the median and $90\%$ symmetric credible intervals of several intrinsic and extrinsic parameters, including the luminosity distance ($D_L$) and the source redshift ($z$), assuming flat $\Lambda$CDM cosmology with the cosmological parameter values inferred by Planck~\cite{Ade:2015xua}. Interestingly, the $(q, \chieff)$ contour in the posterior distribution is reminiscent of the well-known $q-\chieff$ degeneracy of the $(2,2)$ mode~\cite{Cutler:1994ys, Poisson:1995ef, Baird:2012cu}, with the inclusion of HM and precession partially breaking that degeneracy. These results demonstrate the importance of both HM and orbital precession in the source parameter inference of GW151226. 
In Section~\ref{sec:Discussion} we will take a closer look at the impact of including these physical effects in the analysis of GW151226. We will also describe how the combination of smaller $q$, larger $\chieff$, and the presence of precession means that the primary black hole spin must be large.

\vskip 4pt

The log likelihood ($\ln\mathcal{L}$) distribution of the {\default} setup is also shown in Fig.~\ref{fig:corner_plot}. The two-dimensional contour plots in the $\ln\mathcal{L}$ row demonstrate that the low-$q$ and high $\chieff$ region is where the likelihood peaks and is therefore statistically preferred by the data. Notably, the average likelihoods of the {\nohm} and {\nopre} setups are significantly smaller than that of the {\default} setup, highlighting the importance of both HM and orbital precession in shifting the solution towards the low-$q$ and high-$\chieff$ region. Note that the priors used in all of these setups are identical; the absence of posterior density in the low-$q$ and high-$\chieff$ region for the {\nohm} and {\nopre} setups therefore arises due to genuine suppression of the likelihood in that region of parameter space when these effects are neglected.

\begin{table*}[t!]
\begin{center}
\setlength\tabcolsep{3pt}
\begin{tabular}{lcccc}
\specialrule{.1em}{.05em}{.05em} 
\specialrule{.1em}{.05em}{.5em} 
& \textsc{\textsc{Default }} & \textsc{No Higher } & \textsc{No} & \textsc{Default} \\ 
& \textsc{\textsc{(Flat-In-$\chieff$ Prior)}} & \textsc{Multipoles} & \textsc{Precession} & \textsc{\textsc{(isotropic spin Prior)}} \\ 
\specialrule{.1em}{.3em}{.5em} 
Detector-frame chirp mass, $\mathcal{M}^{\rm det}$ [$M_{\odot}$] & $9.75^{+0.13}_{-0.08}$  & $9.73^{+0.07}_{-0.07}$ & $9.73^{+0.06}_{-0.06}$ & $9.73^{+0.08}_{-0.07}$ \\[4pt]
Mass ratio, $q = m_2/m_1  $  & $0.30^{+0.42}_{-0.17}$  & $0.40^{+0.38}_{-0.21}$ & $0.47^{+0.35}_{-0.24}$ & $0.46^{+0.43}_{-0.28}$ \\[4pt]
Primary source mass, $m_1$ [$M_{\odot}$]      & $19.1^{+12.1}_{-7.1}$ & $16.5^{+8.6}_{-4.9}$ & $15.0^{+7.1}_{-3.8}$ & $15.1^{+10.7}_{-4.3}$ \\[4pt]
Secondary source mass, $m_2$  [$M_{\odot}$]   & $5.8^{+2.9}_{-1.7}$ & $6.5^{+2.5}_{-1.8}$ & $7.1^{+2.1}_{-2.0}$ & $7.0^{+2.6}_{-2.4}$ \\[4pt]
Total source mass, $M$ [$M_{\odot}$]          & $24.9^{+10.5}_{-4.3}$ & $23.0^{+6.8}_{-2.6}$ &  $22.0^{+5.1}_{-1.8}$ & $22.2^{+8.3}_{-2.0}$ \\[4pt]
Effective aligned spin, $\chieff$ & $0.30^{+0.25}_{-0.15}$ & $0.26^{+0.21}_{-0.13}$ & $0.24^{+0.20}_{-0.10}$ & $0.23^{+0.22}_{-0.09}$   \\[4pt]
Luminosity distance, $D_L$ [Mpc] & $485^{+117}_{-159}$ & $475^{+129}_{-158}$ & $502^{+153}_{-188}$ & $482^{+131}_{-166}$ \\[4pt]
Source redshift, $z$                 & $0.10^{+0.02}_{-0.03}$ & $0.10^{+0.03}_{-0.03}$ & $0.11^{+0.03}_{-0.04}$ & $0.10^{+0.03}_{-0.03}$ \\[4pt]
Log likelihood, $\ln\mathcal{L}$   & $88.2^{+3.7}_{-4.9}$ & $86.2^{+2.6}_{-4.2}$ & $84.9^{+2.4}_{-4.0}$ & $87.4^{+4.2}_{-4.7}$ \\[4pt]
\specialrule{.1em}{.05em}{.05em} 
\specialrule{.1em}{.05em}{.05em} 
\end{tabular}
\setlength{\belowcaptionskip}{-15pt}
\caption{Median and $90\%$ symmetric credible interval of the source parameters and the log likelihood for the PE setups outlined in Section~\ref{sec:PEsetup}. Posteriors under the flat-in-$\chieff$ prior with different signal models are shown in Fig.~\ref{fig:corner_plot}. A comparison of the posteriors under the flat-in-$\chieff$ versus isotropic spin priors are shown in Fig.~\ref{fig:corner_plot_lvc}.} \label{table:params}
\end{center}
\end{table*}

\vskip 4pt

In Fig.~\ref{fig:corner_plot_lvc}, we present the posterior distributions of the {\textsc{Default (isotropic spin Prior)}} setup in order to determine the effects of the spin prior on our parameter inference. For ease of comparison, we include again the results of the {\default} setup and explicitly label its prior in the legend name for clarity. Due to the isotropic spin prior's suppression of large $|\chieff|$ values, we find that the posterior probability of the high-$\chieff$ region in the {\textsc{Default (isotropic spin Prior)}} case is diminished. Since the high-$\chieff$ region is correlated with the low-$q$ region, we see in the $(q, \chieff)$ contours of Figs.~\ref{fig:corner_plot} and~\ref{fig:corner_plot_lvc} that the low-$q$ region in the isotropic spin case is also relatively suppressed. This finding suggests that, in addition to HM and precession, a spin prior that does not penalize solutions with large values of $|\chieff|$ plays an important role in uncovering that higher likelihood region of parameter space for GW151226. In Table~\ref{table:params} we present the median and $90\%$ symmetric credible intervals of each setup. By visual inspection, we find that the {\textsc{Default (isotropic spin Prior)}} posteriors in Fig.~\ref{fig:corner_plot_lvc} appear similar to the results of~Ref.\cite{Mateu-Lucena:2021siq}, which use the isotropic spin prior. In addition, the median and $90\%$ credible interval as reported in their Table III appears consistent with ours in Table~\ref{table:params}.

\section{Source Discussion} \label{sec:Discussion}

In the previous section, we saw that the posterior from the {\default} setup is shifted towards the low-$q$ and high-$\chieff$ region compared to the {\nohm} and {\nopre} setups. Crucially, the solution in the low-$q$ and high-$\chieff$ region has a larger average log likelihood than the high-$q$ and low-$\chieff$ region. 

\vskip 3pt

\begin{figure}[b]
    \centering
    \includegraphics[width=\linewidth, trim=0 20 0 0 ]{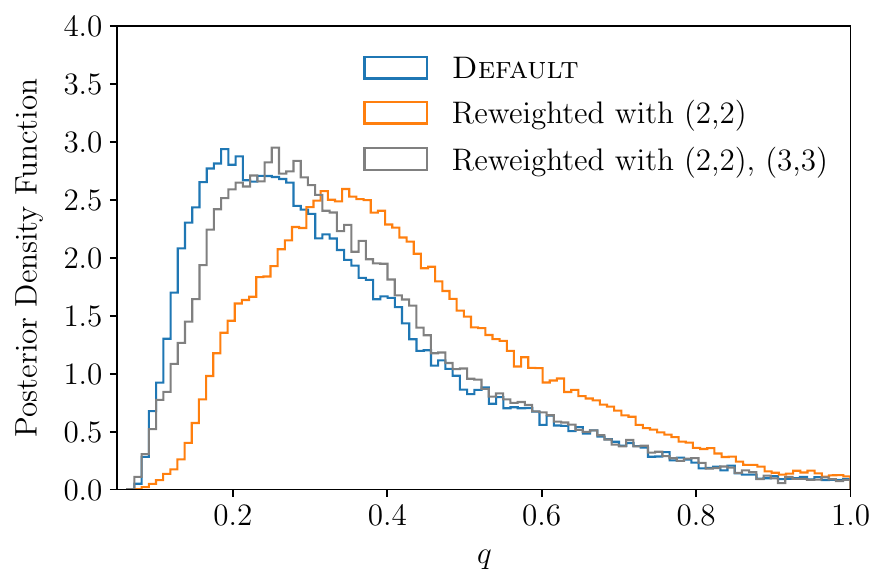}
    \setlength{\belowcaptionskip}{-10pt}
    \caption{Posterior distributions conditioned on $q$ for the {\default} setup and for the scenarios in which the {\default} samples have been reweighted with likelihoods computed using different subsets of the available multipole modes. One reweighting uses only the $(2,2)$ mode in the co-precessing frame, and the other uses the $(2,2)$ and $(3,3)$ modes in the co-precessing frame. In both cases the effects of precession are still included. This figure shows that the co-precessing $(3,3)$ mode, in addition to the dominant $(2,2)$, is important in recovering the GW151226 likelihood peak in the low-$q$ region.}
    \label{fig:Reweight_HM}
\end{figure}

Why is the low-$q$ region relatively suppressed in the {\nohm} and {\nopre} setups? In the former case, this comes from the fact that HM are more strongly excited when the constituent masses are more asymmetric~\cite{Blanchet:2013haa, Kidder:2007rt, Berti:2007fi, Blanchet:2008je}. By omitting HM in the {\nohm} setup, we penalize most heavily the solutions that are most efficient at emitting these modes, which means suppressing extreme mass ratio solutions. Using the mode selection feature of the LALSuite implementation of {\phXPHM}~\cite{Pratten:2020ceb, lalsuite}, we can determine the relative importance of the various multipole modes in our solution. In particular, we achieve this by reweighting the likelihood of the {\default} setup with likelihoods that exclude some or all of the HM, and then compare the resulting posterior distributions~\cite{10.5555/1051451, 10.5555/1571802}. The results are shown in Fig.~\ref{fig:Reweight_HM}, where we find that the $(3,3)$ multipole in the co-precessing frame is important for the shift towards the low-$q$ region. Other HM contribute only marginally. In a more refined analysis, we find that the relative importance of the multipole modes obeys the hierarchy $(2,2) > (3,3) > (2,1) > (4,4) > (3,2)$. We note that Ref.~\cite{Payne:2019wmy} had also analyzed the importance of HM in GW151226 through the likelihood reweighting method described here~\cite{10.5555/1051451, 10.5555/1571802}, though a direct comparison cannot be made since precession was omitted in that work.

\vskip 6pt

\begin{figure*}
    \centering
    \includegraphics[width=0.73\linewidth]{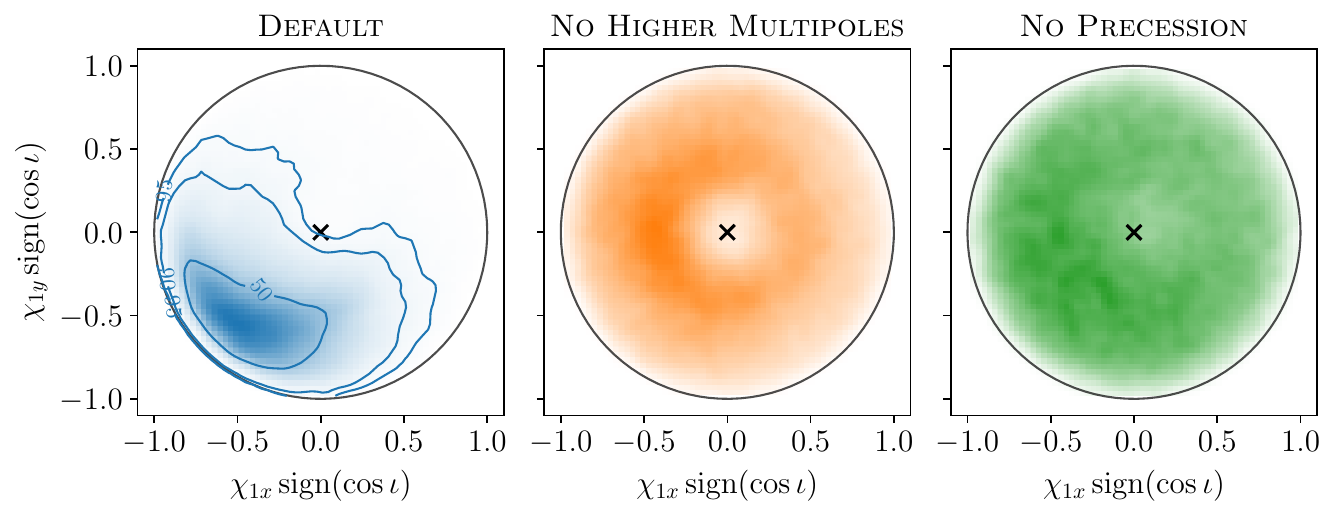}
    \setlength{\belowcaptionskip}{-10pt}
    \caption{Comparison of the in-plane spin posteriors from the various PE setups. We multiply a factor of sign$(\cos \iota)$ on each of the spin components because the figures, especially the left panel, would otherwise display two modes that are similar up to a $180^\circ$ rotation in the azimuthal direction. This follows from an approximate reflection symmetry in the measured orientation of the orbit. For the {\default} setup, we show the $50\%$ and $95\%$ credible regions, finding that zero in-plane spin (black cross) is excluded at the $95\%$ credible level. Due to precessional motion, the position of the {\default} in-plane spin evolves in the azimuthal direction; in this work, we use a reference frequency of $f_{\rm ref} = 120$ Hz. Recall that the {\nopre} setup is designed such that the in-plane spins are ignored in the likelihood evaluation -- the distribution in the right panel therefore corresponds to the in-plane spin prior conditioned on the orbit-aligned spin.}
    \label{fig:in-plane-spin}
\end{figure*}

The suppression of the low-$q$ region in the {\nopre} setup implies that the primary BH has a large in-plane spin. This can be understood from the fact that orbital precession, which is governed by the following equation at leading order~\cite{Kidder:1992fr, Kidder:1995zr, Schmidt:2014iyl}
\begin{equation}
    \frac{d \vec{L}}{d t} = \frac{m_1^2}{r^3} \left[ \left( 2 + \frac{3q}{2} \right) \vec{\chi}_1 + \left( 2 q^2 + \frac{3q}{2} \right) \vec{\chi}_2 \right] \times \vec{L} \, , \label{eqn:precession}
\end{equation}
is mainly driven by the in-plane spin of the heavier BH when $q$ is small. In Fig.~\ref{fig:in-plane-spin}, we plot the posterior distributions of the primary spin's in-plane components for our various PE setups. The secondary spin is unconstrained in all cases, so we do not show it here. We found that the two-dimensional posterior distributions of in-plane spin components, $\chi_{1x}$ and $\chi_{1y}$, exhibited two similar modes that are related through a rotation of $180^\circ$ in the (instantaneous) orbital plane. This is the case because the inclination of the binary with respect to the detector's line-of-sight is only determined up to an approximate symmetry under reflection of the orbital angular momentum across the plane of the orbit. For this reason, in Fig.~\ref{fig:in-plane-spin} we use the coordinates $\chi_{1x} \, \operatorname{sign}(\cos\iota)$ and  $\chi_{1y} \, \operatorname{sign}(\cos\iota)$ to collapse the two degenerate islands into one, where $\iota$ is the inclination angle between the orbital angular momentum and the line-of-sight. Note that the definitions of $\chi_{1x}, \chi_{1y}$ and $\iota$ depend on the reference frequency, $f_{\rm ref}$, which is the radiative frequency of the fundamental harmonic at the time when the dynamical variables (BH spins, orbital inclination and phase) are set to their specified values before evolving according to the precessional motion. In this work, we choose $f_{\rm ref} = \SI{120}{\hertz}$, which is close to the detectors' peak sensitivities.

\vskip 4pt

From Fig.~\ref{fig:in-plane-spin}, we see that the primary BH of the {\default} setup has a large in-plane spin, excluding zero at the $95\%$ credible level. For the {\nopre} setup, we observe that the posterior distribution is approximately uniform over the $\left( \chi_{1x} \, \operatorname{sign}(\cos \iota), \chi_{1y} \, \operatorname{sign}(\cos \iota) \right)$ plane. This is the case because the in-plane spin components do not enter the likelihood evaluation in this setup (see Section~\ref{sec:PEsetup}) and are thus determined only by the prior. The posterior for the {\nopre} setup is therefore a \lib{PyMultiNest} sampling of the in-plane spin prior conditioned on the orbit-aligned component of the spin. We choose this prior to be uniform in $\cos \iota$ and throughout the disk $\chi_{1x}^2 + \chi_{1y}^2 = 1 - \chi_{1z}^2$, with $\chi_{1z}$ being the orbit-aligned spin component on which the in-plane distribution is conditioned. Since the primary in-plane spin posterior of the {\nohm} setup in Fig.~\ref{fig:in-plane-spin} is almost uniformly distributed, we conclude that precession is not well measured in that setup.

 \begin{figure}[b!]
    \centering
    \includegraphics[width=\linewidth]{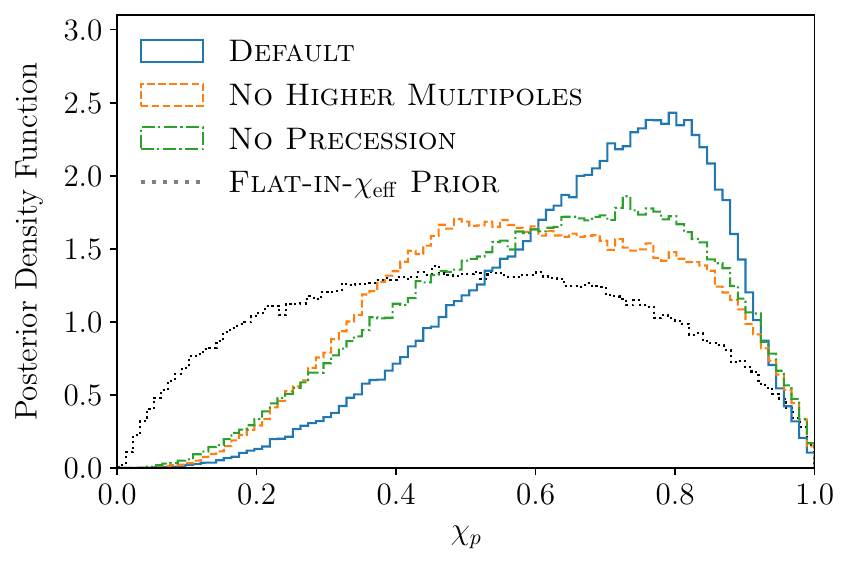}
    \caption{The prior and marginalized posterior distributions on $\chi_p$ for our various PE setups. The distribution of the {\nopre} setup represents the case in which orbital precession is intentionally disabled in the likelihood evaluation, and therefore serves as a useful null comparison for tests of precession in other scenarios.}
    \label{fig:Chip}
\end{figure}

\vskip 4pt

Motivated by the mass-weighted spin combination in (\ref{eqn:precession}), the effective precession parameter~\cite{Schmidt:2014iyl} 
\begin{equation}
    \chi_{p} \equiv \text{max} \left( \big| \vec{\chi}_1 \times  \hat{L} \big| \hskip 2pt , \hskip 4pt \frac{4q^2+3q}{4+3q} \hskip 2pt  \big| \vec{\chi}_2 \times  \hat{L} \big|  \right) \, , \label{eqn:chip}
\end{equation}
is often used as a measure for precession. In Fig.~\ref{fig:Chip}, we plot the $\chi_p$ posteriors from our various PE setups. Recall that the $\chi_p$ distribution of the {\nopre} setup represents the case in which orbital precession is intentionally disabled during likelihood evaluation (see Section~\ref{sec:PEsetup}); it therefore serves as a useful null comparison for tests of precession in other scenarios~\cite{Zackay:2019tzo}. On the one hand, the {\nohm} setup yields a $\chi_p$ posterior very similar to the {\nopre} setup, so we conclude that precession is not well measured in this scenario -- a conclusion that is in agreement with Fig.~\ref{fig:in-plane-spin}. On the other hand, the $\chi_p$ distribution inferred in the {\default} setup manifestly deviates away from the {\nopre} case, indicating that this setup displays signs of precession. However, unlike the in-plane distributions in Fig.~\ref{fig:in-plane-spin}, the $\chi_p$ posteriors do not have an intuitive interpretation. It is therefore difficult to infer the degree to which orbital precession is important in the {\default} setup from Fig.~\ref{fig:Chip} alone.\footnote{As noted in Ref.~\cite{Zackay:2019tzo}, the precise value and shape of any $\chi_p$ distribution should not be overly interpreted, as they depend on the measured values of $q$ and $\chieff$ as well. For instance, although the {\nopre} setup contains no information about precession, its $\chi_p$ posterior in Fig.~\ref{fig:Chip} broadly peaks around $\chi_p \sim 0.8$, instead of $\chi_p \sim 0$, because the Kerr bound $|\vec{\chi}_1| \leq 1$ and  $|\vec{\chi}_2| \leq 1$ correlates $\chi_p$ with the other spin components, with the latter additionally constrained by the data~\cite{LIGOScientific:2018mvr, Zackay:2019tzo}. Similarly, the {\default} setup can be understood to display signs of precession in Fig.~\ref{fig:Chip} not because of the precise shape of its $\chi_p$ distribution, but because of its difference with the $\chi_p$ distribution of the {\nopre} setup.
}

\vskip 4pt

\begin{figure}[t!]
    \centering
    \includegraphics[width=\linewidth, trim=0 20 0 0]{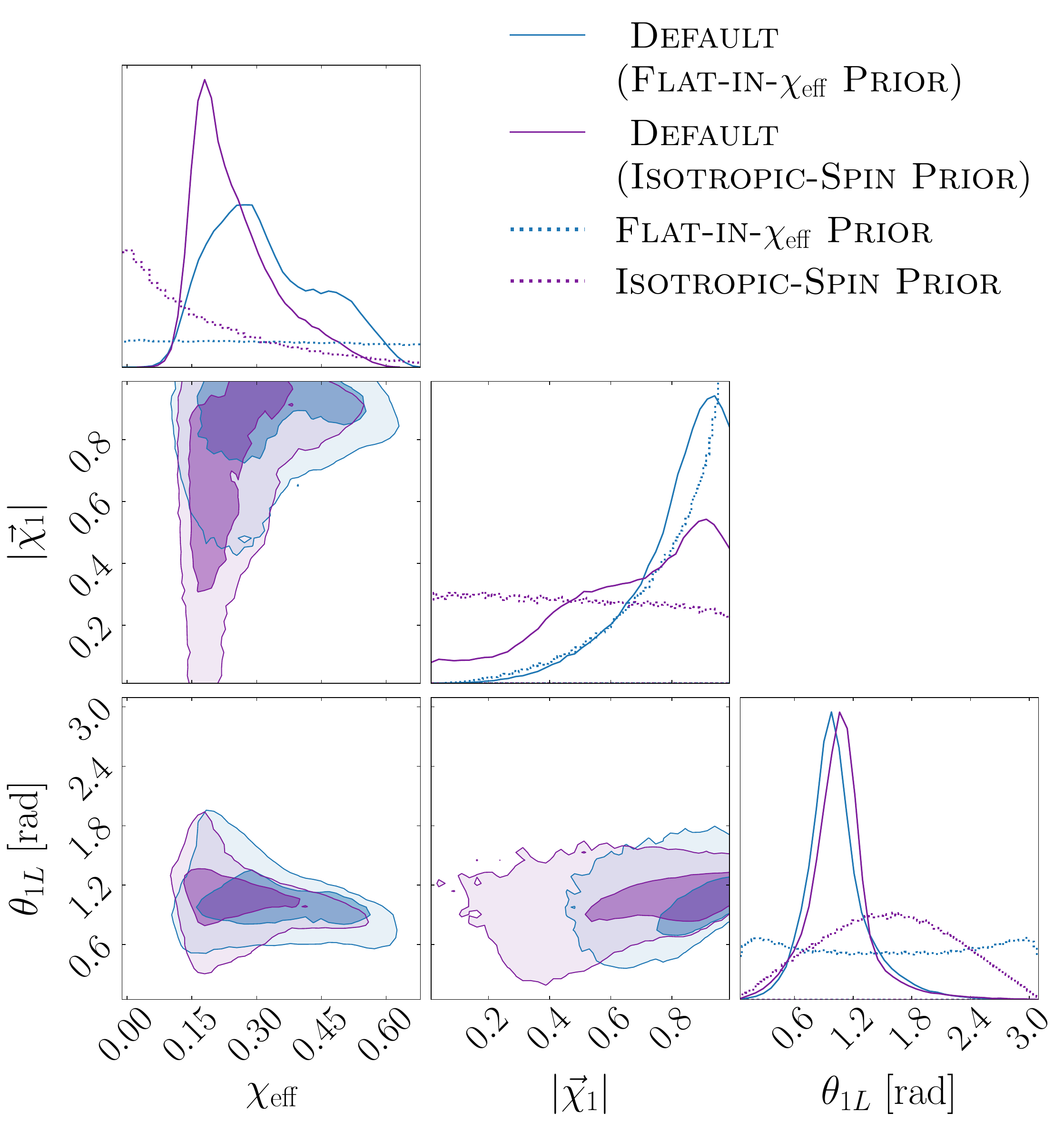}
    \setlength{\belowcaptionskip}{-10pt}
    \caption{Posterior distributions of $\chieff$, primary spin magnitude ($|\vec{\chi}_1|$), and the inclination angle between the primary spin and the orbital angular momentum ($\theta_{1L}$) measured at the reference frequency $f_{\rm ref} = 120$ Hz. Here we show the {\default} posteriors under the flat-in-$\chieff$ prior and the isotropic spin prior, both of which are also in the diagonal subplots (prior distributions are dotted, posteriors are solid).}
    \label{fig:Spin}
\end{figure}

In addition to having a primary BH with non-negligible in-plane spin, the solution of the {\default} setup also has a modestly large value of $\chieff$ (see Fig.~\ref{fig:corner_plot} and Table~\ref{table:params}). This is especially interesting because $\chieff$ is dominated by the orbit-aligned somponent of the primary BH spin in the low-$q$ limit. The primary spin is therefore not only tilted away from the orbital angular momentum, but must also have a large magnitude. In Fig.~\ref{fig:Spin}, we show both the {\default} setup and the \textsc{Default (Isotropic Spin Prior)} setup posterior distributions of $\chieff$, the primary spin magnitude ($|\vec{\chi}_1|$), and the inclination angle between the primary spin and the orbital angular momentum ($\theta_{1L}$) measured at $f_{\rm ref} = \SI{120}{\hertz}$. Comparing the curves in the spin magnitude panels we see that the primary BH indeed spins rapidly, since the posterior weight shifts to large values regardless of the spin prior. The tilt of the primary spin $\theta_{1L} = ({57_{-23}^{+37}})^\circ  $ is not very precisely constrained but rules out alignment and anti-alignment at high confidence under both spin priors. The spin of the secondary BH spin in both setups is virtually unconstrained by the data and is therefore not shown in this paper.

\section{Astrophysical Implications} \label{sec:astro}

Since our flat-in-$\chi_{\rm eff}$ spin prior is arguably less astrophysically motivated than the commonly adopted isotropic spin prior, we do claim that our new solution is the unambiguously favored astrophysical interpretation. It is, however, worth examining if such a system could be produced by current models of stellar populations and dynamics. Because the source properties of our solution are different from those previously inferred in the literature, GW151226 could be a BBH merger with previously unexplored astrophysical interpretations. In fact, the combination of small $q$, moderately large $\chieff$, and non-vanishing precession makes GW151226 an outlier in the known population of merging BBH, as most of the signals detected to date have comparable component masses, small values of $\chieff$, and display no signs of precession~\cite{LIGOScientific:2018mvr, Zackay:2019tzo, Venumadhav:2019tad, Nitz:2019hdf, gwtc2_LIGOScientific:2020ibl, Abbott:2020gyp}. In the following discussion, we briefly explore the astrophysical implications of our solution.

\vskip 4pt 

Asymmetric-mass binary systems are relatively uncommon in the observed population of BBH~\cite{LIGOScientific:2018mvr, gwtc2_LIGOScientific:2020ibl, Zackay:2019tzo, Nitz:2019hdf}. In this respect, our solution for GW151226 is similar to GW190412~\cite{LIGOScientific:2020stg} and GW190814~\cite{Abbott:2020khf}, whose mass ratios of $q = 0.28^{+0.12}_{-0.06}$ and $q = 0.112^{+0.008}_{-0.009}$ make them the two most asymmetric binary systems in the GWTC-2 catalog of LVC detections \cite{gwtc2_LIGOScientific:2020ibl}. The two broad classes of formation channels for merging BBH, which are the canonical isolated binary evolution channel~\cite{1976IAUS...73...75P, 1976IAUS...73...35V, 1993MNRAS.260..675T, Postnov:2014tza} and the dynamical merger channel~\cite{PortegiesZwart:1999nm, 2010MNRAS.407.1946D, Rodriguez:2015oxa, Antonini:2016gqe, Ziosi:2014sra, Antonini:2016gqe, Banerjee:2018pmh, Kumamoto:2018gdg}, typically predict orders of magnitude more comparable-mass BBH than asymmetric-mass systems~\cite{Dominik:2012kk, Eldridge:2016ymr, deMink:2016vkw, Rodriguez:2016kxx, Spera:2018wnw, Neijssel:2019irh}. A measurement of $q$ alone therefore does not immediately inform us about the formation history of GW151226, though it is interesting that GW151226 increases the sample size of asymmetric-mass binary systems, which could provide better statistical constraints on the formation channels mentioned above, including constraints on their branching fractions.

\vskip 4pt

The secondary (source-frame) mass $m_2 = 5.8^{+2.9}_{-1.7}\,\msun$ of our solution has some overlap with the hypothesized lower mass gap of $2.5$--$5\, M_\odot$~\cite{Bailyn:1997xt, Ozel:2010su, Farr:2010tu}. The lower bound of the mass gap is determined by the maximum mass of neutron stars, with the uncertainties governed by the stars' modestly constrained nuclear equation of state~\cite{Lattimer:2012nd, Ozel:2016oaf, Nathanail:2021tay}. On the other hand, the upper bound is empirically inferred from observations of X-ray binaries~\cite{Bailyn:1997xt, Ozel:2010su, Farr:2010tu} and theoretically depends on the details of the explosion and implosion mechanisms of stellar cores~\cite{Janka:2012wk, 2012ApJ...749...91F}. Taking the upper bound to be $5\,M_\odot$, the probability that our solutions falls in this mass gap is $27\%$. Most of the BBH in the GWTC-2 catalog, except GW190814,\footnote{Although the secondary mass of GW190814 was found to be $2.59\, M_\odot$~\cite{Abbott:2020khf}, it remains unclear if it is a heavy neutron star or a light black hole~\cite{Abbott:2020khf, Most:2020bba, Rather:2020lsg, Rather:2021yxo}.} have constituent masses that are well above this gap; in more recent analyses of the O3a data \cite{iasO3a:2022pin, nitz_o3a_3ogc_catalog_2021, gwtc2-1_LIGOScientific:2021usb}, however, as well as in electromagnetic observations \cite{lmg_BH_EMobs_science2019, lmg_confirmingRV_unicorn_close_3msunBH_EMobs_Masuda2021, lmg_unicorn_close_3msunBH_EMobs_Jayasinghe2021}, additional BHs have been detected in this mass range, so GW151226 may contribute to this growing population of lower mass gap events. 

\vskip 4pt

Spin measurements offer valuable probes into the formation mechanisms of merging binaries~\cite{Farr:2017uvj, Rodriguez:2016vmx}. As illustrated in Fig.~\ref{fig:Spin}, the primary spin in GW151226 is not only tilted away from the orbital angular momentum, but also has a large magnitude $|\vec{\chi}_1| = 0.85^{+0.13}_{-0.35}$, and the secondary spin is essentially unconstrained. If GW151226 was formed through the canonical isolated binary evolution channel~\cite{1976IAUS...73...75P, 1976IAUS...73...35V, 1993MNRAS.260..675T, Postnov:2014tza}, the spin measurement in Fig.~\ref{fig:Spin} may suggest that a first-generation BH in this formation channel can attain large spins prior to merger. A highly-spinning primary BH seems theoretically unlikely in this channel because angular momentum transport from the progenitor star's helium core to the hydrogen envelope is typically very efficient during the red supergiant phase, resulting in a remnant core that has little angular momentum remaining when it collapses to a BH \cite{Belczynski:2017gds, Fuller:2019sxi, 2019MNRAS.485.3661F}. Tidal interactions could spin up the helium core, although this would be very inefficient before the common envelope phase~\cite{1976IAUS...73...75P, 1976IAUS...73...35V, 1993MNRAS.260..675T, Postnov:2014tza}, since the binary separation would still be large. That being said, if the binary progenitor stars were born in tight orbits, they could acquire large spins through strong tidal interactions, leading to substantial chemical mixing in both stars~\cite{Mandel:2015qlu, Marchant:2016wow}. This mixing prevents significant expansions of stars, thereby further suppressing angular momentum loss through stellar winds or accretion. Because the spins of field binaries are preferentially aligned with the orbital angular momentum, the observed misalignment between $\vec{\chi}_1$ and $\vec{L}$ would most likely arise from a natal kick imparted by the non-spherical core collapse supernova of the progenitor stars~\cite{Kalogera:1999tq,Willems:2004kk, OShaughnessy:2017eks}. For the chemically-homogeneous evolution channel, the spin misalignment angle is typically small because the orbital velocity is large (due to the small binary separation) compared to the kick velocity~\cite{Mandel:2015qlu}. While this expectation may seem to contradict our $\theta_{1L} = ({57_{-23}^{+37}})^\circ$ finding, that picture largely applies to comparable-mass binary systems $q \gtrsim 0.3$~\cite{Mandel:2015qlu} and could change for binary systems with more asymmetric mass ratios.

\vskip 4pt

If GW151226 was formed through dynamical capture in a dense stellar environment~\cite{PortegiesZwart:1999nm, 2010MNRAS.407.1946D, Rodriguez:2015oxa, Antonini:2016gqe, Ziosi:2014sra, Antonini:2016gqe, Banerjee:2018pmh, Kumamoto:2018gdg}, then the measured spin could be explained as the result of the primary BH being the remnant of a previous merger, making GW151226 a hierarchical merger ~\cite{Gerosa:2017kvu}. In the dynamical capture channel, the BH spins are expected to be randomly and isotropically distributed, which would naturally explain the measured misalignment between $\vec{\chi}_1$ and $\vec{L}$. While the distributions of BH spin magnitudes in dense stellar environments are highly uncertain, in the hierarchical merger scenario~\cite{Doctor:2019ruh, Rodriguez:2019huv}, mergers of first- or higher-generation BHs would form second- or higher-generation BHs with spins that are distributed approximately around $|\vec{\chi}| \approx 0.7$~\cite{Pretorius:2005gq, Scheel:2008rj, Fishbach:2017dwv}, as a result of converting the pre-merger orbital angular momentum to the remnant BH spin. The precise value of the final spin depends on the mass ratio and the spin configuration of the merger constituents~\cite{Hofmann:2016yih, Jimenez-Forteza:2016oae,Deng:2020rnf}. If the primary BH of GW151226 is a merger remnant, the large spin magnitude $|\vec{\chi}_1| = 0.85^{+0.13}_{-0.35}$ could be achieved as long as the spins of its progenitors have modestly large components along the direction of the orbital angular momentum~\cite{Hofmann:2016yih, Jimenez-Forteza:2016oae,Deng:2020rnf}. This spin configuration typically gives rise to a large merger kick, with only a small fraction of remnants expected to be retained by dense stellar environments~\cite{Rodriguez:2019huv}. This small fraction is consistent with the fact that GW151226 is a rare event among the observed population~\cite{LIGOScientific:2018mvr, Zackay:2019tzo, Venumadhav:2019tad, Nitz:2019hdf, gwtc2_LIGOScientific:2020ibl, Abbott:2020gyp}.

\vskip 4pt

Considering the subpopulation of low-mass BBH in the LVC catalogs, let us make an order-of-magnitude estimate of the capture efficiency-- that is, the fraction of merger remnants which are recaptured and merge again-- implied if we take GW151226-like events to be hierarchical mergers. We will make a number of highly simplifying assumptions, so the results should not be over-interpreted.
Neglecting the mass loss due to GW radiation for simplicity, the total mass of a BBH gives a rough approximation of its remnant BH mass. From the GWTC-1 and GWTC-2 catalogs~\cite{LIGOScientific:2018mvr, gwtc2_LIGOScientific:2020ibl}, we find that there are about nine BBH systems whose total masses are distributed within the $90\%$ confidence interval inferred for the primary BH in GW151226, $m_1 = 24.9^{+10.5}_{-4.3} \,M_\odot$. Assuming that all nine were mergers of first-generation BHs in dense stellar environments, the observed ratio of first-generation mergers to GW151226-like mergers would be about $9:2$, where we consider GW190412 to be similar to GW151226 because its mass ratio is also small and its primary spin could be large~\cite{Abbott:2020khf} (i.e., its primary BH could be similarly classified as a merger remnant).\footnote{Although GW190814 has a small mass ratio, it is different from GW151226 and GW190412 because its measured primary spin $|\vec{\chi}_1| \leq 0.07$ is tightly constrained away from the expected value for a merger remnant ( $|\vec{\chi}_1|\approx 0.7$~\cite{gwtc2_LIGOScientific:2020ibl}), and is therefore inconsistent with being a second-generation BH. }

This yields a capture efficiency estimate of $\sim 20\%$.
Since those nine progenitor-mass BBH have $q \gtrsim 0.5$ and their chirp masses are distributed randomly around that of GW151226, the average volume up to which one can observe GW151226 is smaller than those of the first-generation mergers by a factor of about $1$--$1.5$ due to the more asymmetric $q \approx 0.3$ of GW151226~\cite{Fishbach2017, Roulet:2018jbe}. Correcting for this observational bias would increase the estimate above, though not by a significant amount. The naive $\sim 20\%$ estimate appears to be of the same order of magnitude as the capture efficiencies expected from population synthesis models, though the detailed predictions depend sensitively on the initial conditions assumed for the BH and stellar populations in these dense environments~\cite{Rodriguez:2019huv, Liu:2020gif}.

This extremely simplified comparison suggests GW151226-like sources could be attributed to the hierarchical formation channel without ruling out agreement between the observed population and the predictions of population synthesis models, though we note that the estimate above would increase if we relax the optimistic assumption that all nine of the BBH fitting the total mass profile were first-generation mergers in dense clusters. This would also suggest that GW151226 is unlikely to be a hierarchical merger unless a sizeable fraction of detectable events arise from dense stellar clusters. Recent analyses~\cite{Abbott:2020gyp, Zevin:2020gbd} of the O1--O3a binary black hole population suggest that this fraction is indeed sizable, though its value is not very precisely constrained. It will be interesting to see how our estimation above changes when more data is analyzed from future observing runs.

\section{Conclusion} \label{sec:conclusion}

In this paper, we inferred the source parameters of GW151226 with {\phXPHM}~\cite{Pratten:2020ceb}, a quasi-circular BBH model which incorporates orbital precession effects and higher-order modes in the post-Newtonian multipole expansion of the gravitational waveform. Sampling under a prior that is uniform in the detector-frame constituent masses and effective spin, and comparing PE setups with different combinations of precession and HM included in likelihood computations, we find that the posteriors shift towards the low-$q$ and high-$\chieff$ region of parameter space when both HM and precession are included in the signal model (see Fig.~\ref{fig:corner_plot}). This new solution is missed when either of these effects are excluded from the waveform, because the higher-dimensional likelihood manifold in this region of parameter space peaks for primary spins tilted away from the orbital angular momentum (indicating the data's preference for a precessing source, as evident in Fig.~\ref{fig:in-plane-spin}) and the asymmetric masses result in a non-negligible contribution from the $(\ell, |m|) = (3, 3)$ harmonic in the co-precessing frame (see Fig.~\ref{fig:Reweight_HM}). In Fig.~\ref{fig:corner_plot_lvc}, we can see that the the flat-in-$\chieff$ prior is also helpful in uncovering this new likelihood peak, which is difficult to explore under under a prior that suppresses large values of $|\chieff|$.

\vskip 4pt

Since the isotropic spin prior used by LVC and a number of other astrophysical formation channels indeed favor small values of effective spin, this new likelihood peak is perhaps not the most astrophysically probably description of the source. However, our solution raises the intriguing possibility that GW151226 is a different type binary system from what was previously inferred in the literature. This new solution is interesting especially because it possesses source properties that are considered rare among the observed population~\cite{gwtc2_LIGOScientific:2020ibl, Abbott:2020gyp}: its mass ratio is far from unity; its secondary constituent may fall in the lower BH mass gap; and its heavier BH is spinning rapidly at a large tilt with respect to the orbital angular momentum, driving precession and suggesting the possibility of a hierarchical merger. It will be interesting to see how the fraction of similar types of signals in the observed population evolves as the expansion of the catalog reduces statistical errors, and how these detections in future observing runs will shed light on the astrophysical formation mechanisms of merging BBH.



\vskip 4pt

\textit{Note:}\label{note_on_v1} in the first version of this work, our PE results displayed clear bimodality in the GW151226 posteriors. There was a pronounced peak in the low-$q$ region and a separate, broader peak in the high-$q$ region, with the former having a much higher likelihood than the latter. We later found that the inferred posteriors were impacted by inaccuracies in the likelihood computation.
The relative-binning algorithm used to evaluate the likelihood relies on the property that the ratio of waveforms that are nearby in parameter space is a smooth function of the frequency, which can be evaluated at low resolution and interpolated \cite{Zackay:2018qdy}.
For waveforms with higher order modes, however, the ratio is only smooth mode-by-mode, but oscillatory for the full waveform, thereby causing interpolation errors.
The relative-binning algorithm has since been generalized to waveforms with higher modes \cite{Leslie2021}, which we implemented in the present version.

\vskip 2pt

In this updated version, we report PE results from a recently-developed software called \texttt{cogwheel}~\cite{Javier}, which uses optimal sampling coordinates and gains significant speedup from implementing efficient likelihood evaluation with mode-by-mode relative binning. We also increase the time-series Nyquist frequency from 512 Hz to 1024 Hz (this was proposed in Ref.~\cite{Mateu-Lucena:2021siq} as the source of disagreement between our results). We found that the improved accuracy in likelihood evaluation at fixed relative binning resolution plays the dominant role in changing our PE results, in particular moving our posterior under the isotropic spin prior into better agreement with Ref.~\cite{Mateu-Lucena:2021siq}. While our posteriors now are no longer strongly bimodal, we see in Fig.~\ref{fig:corner_plot} that they are still preferentially tilted towards the low-$q$ region (arguably even more so, since the broad high-$q$ mode that we found earlier is now more suppressed than in the previous result). Although our current inferred parameter values in the {\default} setup are slightly different from the low-$q$ mode of the earlier work (since we no longer artificially divide the posterior into two modes), our qualitative conclusions and astrophysical interpretation of GW151226 remain unchanged.

\section*{Acknowledgements}

HSC gratefully acknowledges support from the Rubicon Fellowship awarded by the Netherlands Organisation for Scientific Research (NWO). SO acknowledges support from the National Science Foundation Graduate Research Fellowship Program under Grant No. DGE-2039656. Any opinions, findings, and conclusions or recommendations expressed in this material are those of the authors and do not necessarily reflect the views of the National Science Foundation. LD acknowledges support from the Michael M. Garland startup research grant at the University of California, Berkeley. TV acknowledges support by the National Science Foundation under Grant No. 2012086. BZ is supported by a research grant from the Ruth and Herman Albert Scholarship Program for New Scientists. MZ is supported by NSF grants PHY-1820775 the Canadian Institute for Advanced Research (CIFAR) Program on Gravity and the Extreme Universe and the Simons Foundation Modern Inflationary Cosmology initiative. 

\vskip 4pt

This research has made use of data, software and/or web tools obtained from the Gravitational Wave Open Science Center (https://www.gw-openscience.org/), a service of LIGO Laboratory, the LIGO Scientific Collaboration and the Virgo Collaboration. LIGO Laboratory and Advanced LIGO are funded by the United States National Science Foundation (NSF) as well as the Science and Technology Facilities Council (STFC) of the United Kingdom, the Max-Planck-Society (MPS), and the State of Niedersachsen/Germany for support of the construction of Advanced LIGO and construction and operation of the GEO600 detector. Additional support for Advanced LIGO was provided by the Australian Research Council. Virgo is funded, through the European Gravitational Observatory (EGO), by the French Centre National de Recherche Scientifique (CNRS), the Italian Istituto Nazionale di Fisica Nucleare (INFN) and the Dutch Nikhef, with contributions by institutions from Belgium, Germany, Greece, Hungary, Ireland, Japan, Monaco, Poland, Portugal, Spain.

\bibliographystyle{apsrev4-1}
\bibliography{BoxingDay}

\end{document}